\documentclass[11pt]{article}
\usepackage{latexsym}  
\usepackage{amssymb}
\usepackage{graphicx}
\usepackage{amsmath}

\newcommand{\be}{\begin{equation}}
\newcommand{\ee}{\end{equation}}
\newcommand{\bea}{\begin{eqnarray}}
\newcommand{\eea}{\end{eqnarray}}
\newcommand{\bref}[1]{(\ref{#1})}

\begin{document}
\begin{titlepage}
\begin{flushright}
\end{flushright}
\vspace{4\baselineskip}
\begin{center}
{\Large\bf  Simple mass matrices of neutrinos and quarks consistent with observed mixings and masses}
\end{center}
\vspace{1cm}
\begin{center}
{\large Hiroyuki Nishiura$^{a,}$
\footnote{E-mail:hiroyuki.nishiura@oit.ac.jp}}
and
{\large Takeshi Fukuyama$^{b,}$
\footnote{E-mail:fukuyama@se.ritsumei.ac.jp}}
\end{center}
\vspace{0.2cm}
\begin{center}
${}^{a}$ {\small \it Faculty of Information Science and Technology, 
Osaka Institute of Technology,\\ Hirakata, Osaka 573-0196, Japan}\\[.2cm]

${}^{b} $ {\small \it Research Center for Nuclear Physics (RCNP), Osaka University, Ibaraki, Osaka, 567-0047, Japan}

\vskip 10mm
\end{center}
\vskip 10mm
\begin{abstract}
We propose a simple phenomenological model of quarks-leptons mass
matrices having fundamentally universal symmetry structure.
These mass matrices consist of democratic and semi-democratic mass matrix terms commonly to the neutrino and the quark sectors and have only eight free parameters.
We show that this mass matrix model well reproduces  all the observed values of the MNS lepton and the CKM quark mixing angles, the neutrino mass squared difference ratio, and quark mass ratios, with an excellent agreement. 
The model also predicts $\delta_{CP}^\ell =-94^\circ$ for the leptonic $CP$ 
violating phase and $\langle m\rangle\simeq 0.0073$ eV for the 
effective Majorana neutrino mass.
\end{abstract}
PCAC numbers:  
  12.15.Ff, 
  14.60.Pq,  
  12.60.-i, 
\end{titlepage}
The Cabbibo-Kobayashi-Maskawa (CKM) quark mixing matrix is almost diagonal, whereas the Maki-Nakagawa-Sakata (MNS) lepton mixing matrix~\cite{MNS}~\cite{Pontecorvo} is almost maximally mixed. 
The origin of nearly maximally lepton mixing has been  investigated from the point of neutrino mass matrix structure using a $\mu-\tau$ symmetry \cite{F-N} which predicts $\theta_{23}=\pi/4$ and $\theta_{13}=0$. The recent finding~\cite{theta13}~\cite{Daya-Bay}~\cite{RENO}~\cite{Daya-Bay2} of a relatively large $\theta_{13}$ and their global best fits \cite{Schwetz} \cite{Valle} \cite{Cappozzi} forces us to consider its origin and model extensions.   
The rather large $\theta_{13}$  also opens the possibility of CP violation in lepton sector. 
There are theoretical discussions on the CP violating phase. 
In the previous paper \cite{N-F}, we proposed a simple two parameter complex mass matrix for Majorana neutrinos incorporating CP violating phase, although the model predicted small CP violating effect.

In this paper, we propose new phenomenological complex mass matrices $M_\nu$,  $M_u$, and  $M_d$ for Majorana neutrinos and up-type and down-type quarks with  incorporating almost maximal leptonic  CP violating phase and incorporating the quark mass matrices with similar structure as neutrino mass matrix too.
The model of the present paper has only eight free parameters to describe mixings and mass ratios for quarks and neutrinos.
The mass matrices are assumed to take the following forms in the base that charged lepton mass matrix $M_e$ is diagonal:
\be
M_\nu=k_\nu \left[\left(
\begin{array}{ccc}
1 & 1 & 1\\
1 & 1 & 1\\
1 & 1 & 1
\end{array}
\right)
+
\rho_\nu e^{i\frac{\pi}{4}} 
\left(
\begin{array}{ccc}
0 & 0 & 0\\
0 & e^{-i\varphi_\nu} & e^{i\varphi_\nu}\\
0 & e^{i\varphi_\nu} & e^{-i\varphi_\nu} 
\end{array}
\right)
+
z_\nu \left(
\begin{array}{ccc}
0 & 0 & 1\\
0 & 0 & 0\\
1 & 0 & 0
\end{array}
\right)
\right],\label{mass_matrix_neutrino}
\ee

\be
M_u=k_u P^\dagger \left[\left(
\begin{array}{ccc}
1 & 1 & 1\\
1 & 1 & 1\\
1 & 1 & 1
\end{array}
\right)
+
\rho_u e^{i\frac{\pi}{4}} 
\left(
\begin{array}{ccc}
0 & 0 & 0\\
0 & 1 & 1\\
0 & 1 & 1 
\end{array}
\right)
+
z_u \left(
\begin{array}{ccc}
1 & 0 & 0\\
0 & 1 & 0\\
0 & 0 & 1
\end{array}
\right)
\right]P, \label{mass_matrix_up}
\ee

\be
M_d=k_d \left[\left(
\begin{array}{ccc}
1 & 1 & 1\\
1 & 1 & 1\\
1 & 1 & 1
\end{array}
\right)
+
\rho_d e^{i\frac{\pi}{4}} 
\left(
\begin{array}{ccc}
0 & 0 & 0\\
0 & 1 & 1\\
0 & 1 & 1 
\end{array}
\right)
+
z_d \left(
\begin{array}{ccc}
0 & 0 & 1\\
0 & 0 & 0\\
1 & 0 & 0
\end{array}
\right)
\right]. \label{mass_matrix_down}
\ee
Here, $\rho_f$, $z_f$ are real parameters, and $\varphi_\nu$ is a phase parameter which appears in the neutrino mass matrix. 
$P$ is a diagonal phase matrix defined by  

\be
P=\left(
\begin{array}{ccc}
e^{i\phi} & 0 & 0\\
0 & 1 & 0\\
0 & 0 & 1\\
\end{array}
\right). 
\ee

The mass matrices \bref{mass_matrix_neutrino}-\bref{mass_matrix_down} consist of three terms. The first term is a democratic mass matrix, while the  second term is also semi democratic only in the  second and third generations. The third term is a small correction term.
All these terms are symmetric (up to the small phase $\phi$ for $M_u$) and have the universal $2-3$ symmetry \cite{F-N} \cite{F-K-N} (up to the small $z_\nu$ and $z_d$ corrections). 

Since we are interested only in the mass ratios and mixings, we neglect the overall constant $k_f$ in the following discussions. Then, this model has only eight free parameters to explain sixteen observables of four up and down quark mass ratios, three quark mixing angles and one  CP violation phase,
two neutrino mass ratios, three lepton mixing angles and three CP violation phases (one Dirac and two Majorana phases).

From a theoretical point of view, let us give some physical implications on our models.
Phenomenological mass matrix model like this are frequently situated as the preliminary step towards the final systematic interpretation of mass matrices like GUT.
Let us consider some relation of the above mass matrices with the most predictive GUT model, the minimal SO(10) model which is composed of two Higgs Yukawa couplings with $10$ and $\overline{126}$-plets Higgs feilds \cite{Babu} \cite{Fuku1}.
\begin{eqnarray}
\label{massmatrix}
   M_u &=& c_{10} M_{10} + c_{126} M_{126},~~ M_d = M_{10} +   M_{126}   \nonumber \\
   M_D &=& c_{10} M_{10} -3 c_{126} M_{126},~~ M_e = M_{10} -3     M_{126} \\
   M_L &=& c_L M_{126},~~M_R = c_R M_{126}   \nonumber\;. 
\end{eqnarray} 
Here $M_D$, $M_L$, and $M_R$ denote the mass matrices 
of Dirac neutrino, left-handed Majorana, 
and  right-handed Majorana neutrino, respectively with complex constants $c_i$. 
They are all symmetric mass matrices because of the property of $10$ and $\overline{126}$-plets representations. 
From the renormalizability, we may add $120$-plet Higgs but it gives antisymmetric matrix. 
In this case mass matrices are considered as Hermitian with real Yukawa coupling of 10-and
$\overline{\mbox{126}}$-plets Higgs and pure imaginary Yukawa coupling in 120-plet Higgs.
So it is very important whether the mass matrices are symmetric or not.
As is well known, $m_b=m_\tau$ is well valid at $\Lambda_{GUT}\approx 10^{16}$ GeV if we consider the SUSY threshold correction \cite{Hall}. This indicates that
on the $(3,3)$ component of mass matrices, $M_{10}$ dominates over $M_{126}$.
Of course Eq.~(\ref{massmatrix}) works at $\Lambda_{GUT}$ scale and some deviations occur especially in quark masses and mixings and lepton masses at low energy scale $\Lambda_{EW}$.
The renormalization group equation (RGE) effect crucially depends on the Higgs field vevs at the intermediate energy scales. It was recently found after elaborate scan that $\Lambda_R=\Lambda_{GUT}$ still remains valid within $2\sigma$ \cite{mimura}, where $\Lambda_R$ is the seesaw scale ($M_R$ scale). This is very important since $\Lambda_R\ll \Lambda_{GUT}$ spoils the gauge coupling unification at $\Lambda_{GUT}$.
Recently Luo and Xing discussed $\mu-\tau$ symmetry in neutrino mass matrix and its observed small breaking by the RGE from $\Lambda_R=10^{14}$ GeV \cite{Luo}. Also, starting from a degenerate neutrino mass at $\Lambda_R$, Babu, Ma, and Valle obtained a $\mu-\tau$ symmetry up to phase at $\Lambda_{EW}$ by RGE \cite{Ma}. 

It should be emphasized that RGE has quite different aspects in SUSY or Non-SUSY.
In SUSY, for instance, 4-point effective operator of type I seesaw suffers only loop corrections of wave functions of Higgs and leptons (nonrenormalization theorem).
Such RGE effects for lepton come mainly from gauge loop due to smallness of Yukawa coupling, which is flavor blind and MNS and mass ratios remains constant.
As for quark mass ratios and CKM, they suffer some changes in the third family.
As was said, our model in this paper is purely phenomenological but very predictive, and gives some hints on the above mentioned relations between the phenomenological models and GUT.

The Majorana neutrino mass matrix $M_\nu$ is diagonalized by unitary matrix $U$ as follows
\be
M_\nu=U\left(
\begin{array}{ccc}
m_{\nu 1}&0&0\\
0&m_{\nu 2}&0\\
0&0&m_{\nu 3}
\end{array}
\right)U^T.
\ee
Here $m_{\nu i}$ are neutrino masses.
The $U$ is the MNS lepton mixing matrix itself since we assume that the mass matrix for the charged leptons is diagonal in the present model. The MNS lepton mixing matrix $U$ is represented in the standard form,
\bea
U_{std}&=& \mbox{diag} (e^{i \alpha_1^e}, e^{i \alpha_2^e}, e^{i \alpha_3}) U\\ \nonumber
&=&\left(
\begin{array}{ccc}
c_{12}c_{13}&s_{12}c_{13}&s_{13}e^{-i\delta_\nu }\\
-s_{12}c_{23}-c_{12}s_{13}s_{23}e^{i\delta_\nu }&
c_{12}c_{23}-s_{12}s_{13}s_{23}e^{i\delta_\nu}&c_{13}s_{23}\\
s_{12}s_{23}-c_{12}s_{13}c_{23}e^{i\delta_\nu }&
-c_{12}s_{23}-s_{12}s_{13}c_{23}e^{i\delta_\nu }&c_{13}c_{23}\\
\end{array}
\right)
\left(
\begin{array}{ccc}
1 & 0 & 0\\
0 & e^{i\beta} & 0\\
0 & 0 & e^{i\gamma}\\
\end{array}
\right). \label{mixing}
\eea
Here, $e^{i \alpha_i^e}$ comes from the rephasing in the charged lepton fields to make the choice of phase convention, and $c_{ij} \equiv \cos\theta_{ij}$ and $s_{ij} \equiv   \sin\theta_{ij}$ with $\theta_{ij}$ being lepton mixing angles. 
The $\delta_\nu $ is the Dirac CP violating phase and $\beta $ and $\gamma $ are the Majorana CP violating phases.

In the present model, the MNS mixing parameters ($\sin^2 2 \theta_{12}$, 
$\sin^2 2 \theta_{23}$, and $\sin^2 2 \theta_{13}$) and the neutrino mass squared difference ratio ($R_\nu \equiv {\Delta m_{21}^2}/{\Delta m_{32}^2}$) are functions of only three free parameters $\rho_\nu$, $\varphi_\nu$,  and $z_\nu$ defined in (\ref{mass_matrix_neutrino}).  
The observed values of the MNS mixing parameters and $R_{\nu}$ are \cite{PDG14}
\be
\begin{array}{l}
\sin^2 2\theta_{12} = 0.846 \pm 0.021, \\
\sin^2 2\theta_{13} = 0.093 \pm 0.008, \\
\sin^2 2\theta_{23} =0.999^{+0.001}_{-0.018} ,  \label{exptdata1}
\end{array}
\ee
\be
R_{\nu} \equiv \frac{\Delta m_{21}^2}{\Delta m_{32}^2}
=\frac{m_{\nu2}^2 -m_{\nu1}^2}{m_{\nu3}^2 -m_{\nu2}^2}
 = (3.09\pm0.015 ) \times 10^{-2} .\label{exptdata2}
\ee
By fine tuning three parameters  $\rho_\nu$, $\varphi_\nu$, and $z_\nu$, our mass matrix model \bref{mass_matrix_neutrino} well reproduces all the observed three MNS mixing angles and the neutrino 
mass squared difference ratio, \bref{exptdata1} and \bref{exptdata2}, for the case of normal neutrino mass hierarchy  if we take values of  $\rho_\nu$,  $\varphi_\nu$, and $z_\nu$ as 
\be
\rho_\nu=4.06, \quad \varphi_\nu=170.86^\circ, \quad z_\nu=-0.42. \label{abvalue}
\ee
In Fig.~1,  we draw the contour curves of the observed values in \bref{exptdata1} and \bref{exptdata2}  
of the MNS mixing parameters ($\sin^2 2 \theta_{12}$, 
$\sin^2 2 \theta_{23}$, and $\sin^2 2 \theta_{13}$) and the neutrino mass squared difference ratio $R_\nu$ in the $(\varphi_\nu, \rho_\nu)$ parameter plane by fixing $z_\nu$.  
We find that the parameter set around $(\varphi_\nu, \rho_\nu)=(170.86, 4.06)$ and $z_\nu=-0.42$  
indicated by a star ($\star$) in Fig.~1 is consistent with all the observed values.

We now present the predictions in the neutrino sector of our model 
by taking the values for the free parameters \bref{abvalue} as follows:
\bea
\sin^2 2\theta_{12} = 0.867, \\
\sin^2 2\theta_{13} = 0.0901, \\
\sin^2 2\theta_{23} = 0.9998, \\
R_\nu = 0.0293,\\
\delta_\nu = -94.4^\circ, \label{leptonic_CP_phase}
\eea
which are consistent with the observed values \bref{exptdata1} and 
\bref{exptdata2}. It should be noted that the present model predict 
$\delta_\nu\simeq -\frac{\pi}{2}$ for the leptonic  CP violation phase, 
which is very near to the global best fit value \cite{Cappozzi}.

The following absolute neutrino masses and the effective Majorana neutrino mass \cite{Doi1981} 
$\langle m \rangle$ 
in the neutrinoless double beta decay are also predicted, for the parameters given by \bref{abvalue}: 
\be
m_{\nu 1} \simeq 0.0067\ {\rm eV}, \ \ m_{\nu 2} \simeq 0.011 \ {\rm eV}, 
\ \ m_{\nu 3} \simeq 0.051 \ {\rm eV}  ,
\ee
\be
\langle m \rangle =\left|m_{\nu 1} (U_{e1})^2 +m_{\nu 2} 
(U_{e2})^2 +m_{\nu 3} (U_{e3})^2\right| 
\simeq 7.3 \times 10^{-3}\ {\rm eV}.
\ee
Here we use the input value \cite{PDG14}  $\Delta m^2_{32}\simeq 0.00244$ eV$^2$ to predict the absolute neutrino masses.

Now we discuss the CKM quark mixing and the quark mass ratios.
The CKM quark mixing matrix ($\equiv V$) is represented by
\be
V=U_u^\dagger P U_d.
\ee
Here  $U_u$ and $U_d$ are, respectively, unitary matrices which diagonalize the up and down quark  mass matrices $M_u$ and $M_u$ given by \bref{mass_matrix_up} and \bref{mass_matrix_down} as 
\be
M_u=P^\dagger U_u\left(
\begin{array}{ccc}
m_{u}&0&0\\
0&m_{c}&0\\
0&0&m_{t}
\end{array}
\right)U_u^\dagger P,
\ee
\be
M_d=U_d\left(
\begin{array}{ccc}
m_{d}&0&0\\
0&m_{s}&0\\
0&0&m_{b}
\end{array}
\right)U_d^\dagger.
\ee
Here $m_u$, $m_c$, and $m_t$ are up quark  masses, and  $m_d$, $m_s$, and $m_b$ are down quark  masses. These mass eigenvalues depend on the parameters $\rho_f$ and $z_f \ \ (f=u,d)$, but do not on  $\phi$.

From the observed up- and down- quark  mass ratios at $\mu= m_Z$ \cite{q-mass} ,
\be
r^u_{12} \equiv \sqrt{\frac{m_u}{m_c} }= 0.045^{+0.013}_{-0.010}, \ \ \ \ 
r^u_{23} \equiv \sqrt{\frac{m_c}{m_t} } =0.060 \pm 0.005,
\ee
\be
r^d_{12} \equiv \frac{m_d}{m_s} =  0.053^{+0.005}_{-0.003}, \ \ \ \ 
r^d_{23} \equiv \frac{m_s}{m_b} =  0.019 \pm 0.006 ,
\ee
we obtain the best fit values for the parameters $\rho_f$ and $z_f \ \ (f=u,d)$ as
\be
\rho_u = -0.0156, \ \ \ \ 
z_u =0.000021, \label{rzvalue1}
\ee
\be
\rho_d = -0.0690, \ \ \ \ 
z_d = 0.0259.\label{rzvalue2}
\ee
By taking the best fit values \bref{rzvalue1} and \bref{rzvalue2}, we predict   
\be
r^u_{12} =0.045, \ \ \ \
r^u_{23} =0.059, \ \ \ \
r^d_{12} =0.053, \ \ \ \
r^d_{23} =0.022 .
\ee

Since  the parameters $\rho_f$ and $z_f \ \ (f=u,d)$ are fixed by the quark mass ratios,  the CKM mixing matrix is depend on only one remaining parameter $\phi$.  In Fig~2, we draw the curves of the absolute values of the CKM mixing matrix elements $|V_{us}|$, $|V_{cb}|$, $|V_{ub}|$, and $|V_{td}|$ as functions of $\phi$. We find that 
all the observed values \cite{PDG14} \cite{UTfit} given by 
\be 
|V_{us}|=0.22536 \pm 0.00061, \ \ \ \
|V_{cb}|=0.0414 \pm 0.0012 , \ \ \ \ 
\ee 
\be
|V_{ub}|=0.00355 \pm 0.00015,  \ \ \ \
|V_{td}|=0.00886_{-0.00032} ^{+ 0.00033}, 
\ee
\be
\delta_{\rm CP}^q=69.4^\circ \pm 3.4^\circ, 
\ee
are consistent with  the parameter $\phi$ by taking 
\be
\phi=-6.3^\circ.  \label{phivalue}
\ee
Now let us present the predictions of the CKM matrix elements and the  CP violating phase $\delta_{\rm CP}^q$  in the quark sector from our model. 
The best fit parameter values  \bref{phivalue} with \bref{rzvalue1} and \bref{rzvalue2} predicts        
\be
|V_{us}|= 0.2253 , \ \ \  |V_{cb}|= 0.04196 ,  \ \ \ |V_{ub}|= 0.003664 , \\
\ \ \ |V_{td}|= 0.009206    , \ \ \  \delta_{\rm CP}^q=75.1^\circ   .
\ee

Finally let us comment on the difference of the structure of the neutrino mass matrix $M_\nu$ between the present model and the previous model \cite{N-F}. 
In the previous paper \cite{N-F}, we confined ourself in the neutrino sector in order to construct a phenomenological mass matrix model  as simple as possible, and we proposed a simple two parameter complex mass matrix for Majorana neutrinos which consists of a pseudo democratic mass
matrix term and the term which is semi democratic only in the second and third generations. However, the pseudo democratic mass matrix term is not completely democratic because  we assumed that (1,1) and (3,3) elements of the pseudo democratic mass matrix term are  $-1$ while the other elements are $1$. Because of this structure, the previous model well describes all the observed lepton mixings and the neutrino mass squared difference ratio $R_\nu$ by only two free parameters, although the model  predicted small CP violating effect \cite{N-F}.  In the present paper,  we revised the structure of $M_\nu$ so as to have complete democratic mass matrix term in $M_\nu$ as is shown in the first term in \bref{mass_matrix_neutrino}, with which we succeeded  in predicting  large CP violating effect in the lepton sector as is shown in \bref{leptonic_CP_phase}.
However, it is found that we need to introduce another $z_\nu$ term  in \bref{mass_matrix_neutrino} in order to reproduce the observed lepton mixing angles and the neutrino mass squared difference ratio $R_\nu$. 
In Fig.~3, we present $z_\nu$ dependence of the lepton mixing angles, the leptonic Dirac  CP violating phase and $R_\nu$ of the present model with taking $\rho_\nu=4.06$ and $\varphi_\nu=170.86^\circ$. As is depicted in  Fig.~3,  the leptonic CP violating effect is large irrespectively of the value of $z_\nu$ , and we succeeded phenomenologically in constructing a model which is consistent with all the observed lepton mixing angles and $R_\nu$ by choosing  $z_\nu=-0.42$.

In conclusion, we have proposed a simple phenomenological model of quarks-leptons mass
matrices having fundamentally universal symmetry structure with small number of free parameters. 
These mass matrices consist of three mass matrix terms, namely, democratic term, semi-democratic term only in the second and third generations, and corrections from those terms. 
These mass matrices are all symmetric (up to the small phase of $\phi$ for $M_u$) 
and have the universal $2-3$ symmetry (up to the small $z_\nu$ and $z_u$ corrections). 
This mass matrix model reproduces all the observed values of the neutrino mixing,  quark mixing, and neutrino and  quark masses consistently. The model also 
predicts $\delta_{CP}^\ell =-94^\circ$ for the unknown leptonic CP 
violating phase and $\langle m\rangle\simeq 0.0073$ eV for the 
effective Majorana neutrino mass. 
Thus our model is not only predictive but also has the clear symmetry
patterns, which can be some bridge between phenomenological models and
more fundamental models like GUT.

\section*{Acknowledgements}
The work of T.F.\ is supported in part by the Grant-in-Aid for Science Research
from the Ministry of Education, Science and Culture of Japan
(No.~26247036).

\newpage
\begin{figure}[ht]
\begin{center}
\includegraphics[height=.4\textheight]{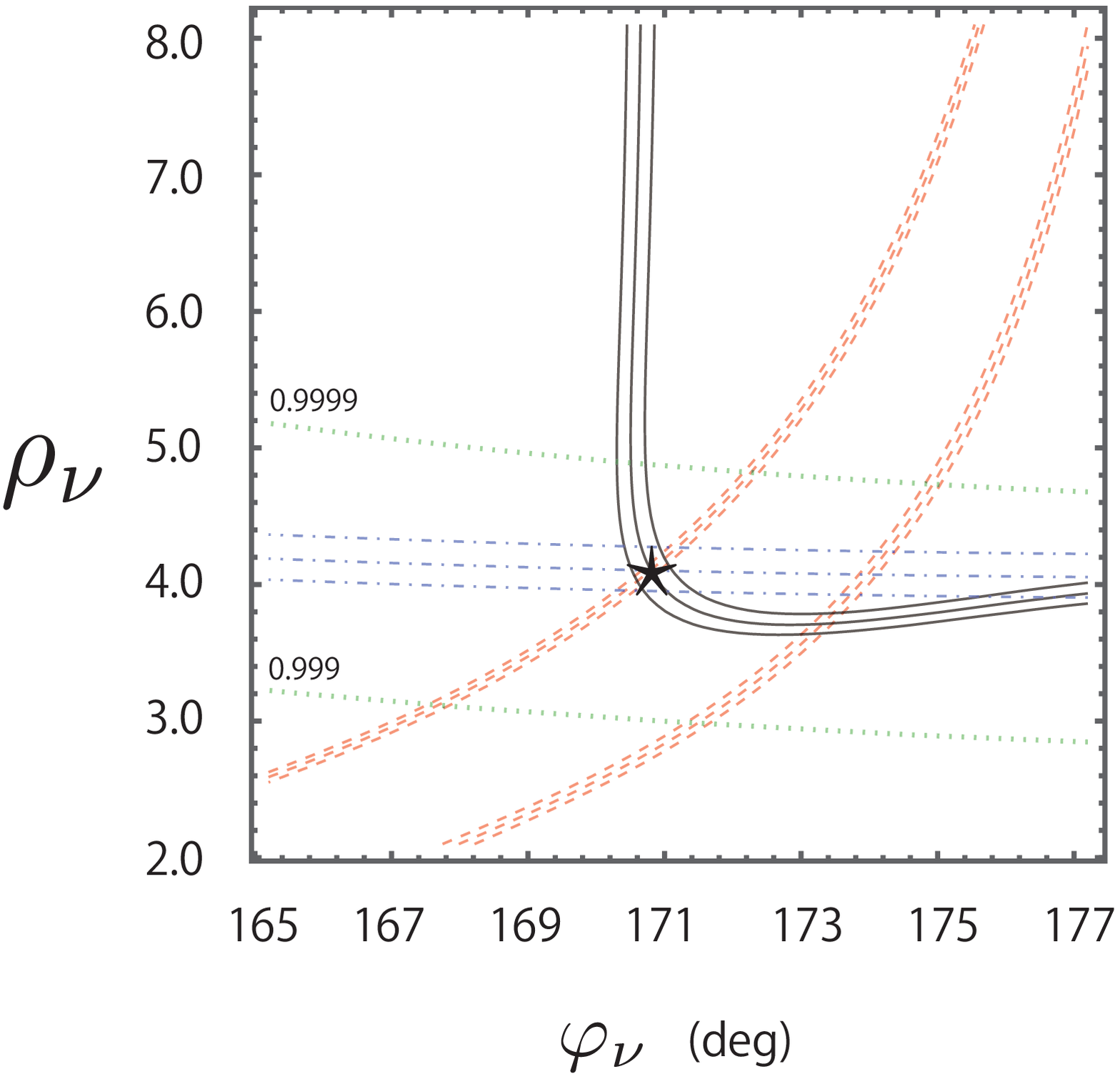}
\end{center}  
  \caption{Contour plots in the $(\varphi_\nu, \rho_\nu)$ parameter plane for the lepton mixing parameters 
$\sin^2 2\theta_{12}$(dashed), 
$\sin^2 2\theta_{23}$(dotted), $\sin^2 2\theta_{13}$(dotdashed), and the neutrino 
mass squared difference ratio $R_\nu$(solid).  Contour curves for the observed (center, upper, and lower) values given in \bref{exptdata1} and \bref{exptdata2} are drawn in the $(\varphi_\nu, \rho_\nu)$ parameter plane with taking $z_\nu=-0.42$ . As for the $\sin^2 2\theta_{23}$, contour curves for $\sin^2 2\theta_{23}=  0.999$ and  $0.9999$ are drawn. 
The parameter set around $(\varphi_\nu, \rho_\nu)=(170.86, 4.06)$   
indicated by a star ($\star$) is consistent with all the observed values.
}\label{fig1}
\end{figure}
\newpage
\begin{figure}[ht]
\begin{center}
\includegraphics[height=0.4\textheight]{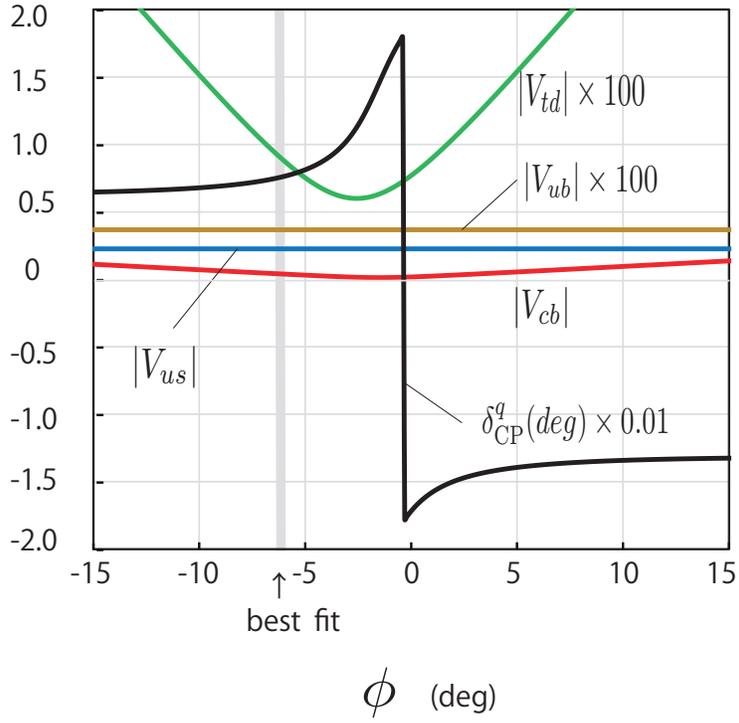}
\end{center}  
  \caption{$\phi$ dependence of the absolute values of the CKM mixing matrix elements $|V_{us}|$, $|V_{cb}|$, $|V_{ub}|$, and $|V_{td}|$, and $\delta_{\rm CP}^q$. 
We have drawn curves of them as functions of $\phi$ 
with taking $\rho_u = -0.0156$, 
$z_u =0.000021$, $\rho_d = -0.0690$, 
and $z_d = 0.0259$.
}\label{fig2}
\end{figure}
\newpage
\begin{figure}[ht]
\begin{center}
\includegraphics[height=.4\textheight]{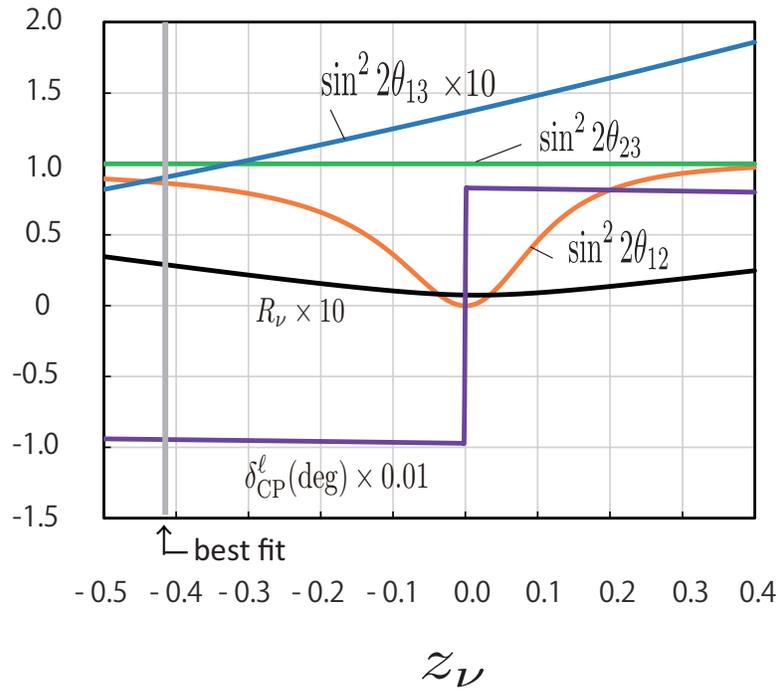}
\end{center}  
  \caption{$z_\nu$ dependence of the lepton mixing parameters 
$\sin^2 2\theta_{12}$, 
$\sin^2 2\theta_{23}$, $\sin^2 2\theta_{13}$, and the neutrino 
mass squared difference ratio $R_\nu$.  The curves are drawn with taking $\rho_\nu=4.06$ and $\varphi_\nu=170.86^\circ$.
}\label{fig3}
\end{figure}

\end{document}